\documentclass[11pt,english,amssymb,nofootinbib,superscriptaddress]{revtex4}
\usepackage{lmodern}

\usepackage[T1]{fontenc}
\usepackage[latin9]{inputenc}
\usepackage{color}
\usepackage{float}
\usepackage{amsmath}
\usepackage{amssymb}
\usepackage{graphicx}
\usepackage{esint}
\usepackage{babel}
\usepackage[unicode=true,pdfusetitle,
 bookmarks=true,bookmarksnumbered=true,bookmarksopen=false,
 breaklinks=false,pdfborder={0 0 1},backref=false,colorlinks=true]
 {hyperref}

\makeatletter

\@ifundefined{textcolor}{}
{%
 \definecolor{BLACK}{gray}{0}
 \definecolor{WHITE}{gray}{1}
 \definecolor{RED}{rgb}{1,0,0}
 \definecolor{GREEN}{rgb}{0,1,0}
 \definecolor{BLUE}{rgb}{0,0,1}
 \definecolor{CYAN}{cmyk}{1,0,0,0}
 \definecolor{MAGENTA}{cmyk}{0,1,0,0}
 \definecolor{YELLOW}{cmyk}{0,0,1,0}
}

\@ifundefined{definecolor}{\usepackage{color}}{}
\usepackage{babel}

\usepackage{latexsym}\usepackage{bm}

\makeatother

\usepackage{babel}
\begin{document}

\title{Minimal Extension of Einstein's Theory: The Quartic Gravity }

\author{Atalay Karasu, Esin Kenar, Bayram Tekin}

\email{\{karasu,aesin,btekin\}@metu.edu.tr}

\selectlanguage{english}%

\affiliation{Department of Physics,\\
 Middle East Technical University, 06800 Ankara, Turkey}

\date{\today}
\begin{abstract}
We study structure of solutions  of the recently constructed minimal extensions of Einstein's gravity in four dimensions at the quartic curvature level. The extended higher derivative theory, just like Einstein's gravity,  has only a massless spin-two graviton about its unique maximally symmetric vacuum. The extended theory does not admit the Schwarzschild or Kerr metrics as  exact solutions, hence there is no issue of Schwarzschild  type singularity but, approximately, outside a source, spherically symmetric metric with the correct Newtonian limit is recovered. We also show that for all Einstein space-times,  square of the Riemann tensor (the Kretschmann scalar or the Gauss-Bonnet invariant) obeys a non-linear scalar Klein-Gordon equation.
\end{abstract}
\maketitle

\section{Introduction}

 Einstein's gravity, as an effective theory, is expected to be modified at small and large distances. Both regimes require different types of modifications. At small distances, which is the subject of this work, experience from quantum field theory and computations in the microscopic theories, such as string theory, suggest that the purely gravitational sector of the  low energy quantum gravity action will be of the form 
\begin{equation}
I=\int d^{n}x \sqrt{-g}\left\{ \frac{1}{2 \kappa_0}\left(R-2\Lambda_{0}\right)+\sum_{i=2}^{\infty}a_{i}\left({\rm Riem,{\rm Ric},{\rm R},\nabla{\rm Riem},\dots}\right)^{i}\right\} ,
\label{action1}
\end{equation}
where $a_i$ are dimensionful quantities computed in the microscopic theory, a virtually impossible task beyond a few small orders as they arise at multi-loop levels. In the absence of a quick way to find the proper actions of low energy quantum gravity such as (\ref{action1}), we might take a more pragmatic way and search for actions that carry the properties of Einstein's gravity, which in this language is the lowest order, best tested, effective quantum gravity theory. For this purpose, let us recapitulate the properties of Einstein's gravity:  It has some obvious virtues, such as diffeomorphism invariance,  uniqueness of the vacuum  (flat space in the absence of the cosmological constant, (anti)-de Sitter in the presence of it) and a massless  unitary spin-2 graviton about its vacuum.  Even though, diffeomorphism invariance is 
easily inherited by the extended gravity actions, the latter two properties, uniqueness of the vacuum and the massless graviton as the only excitation is usually lost.  For example if one considers the quadratic gravity in four dimensions \cite{Stelle1, Stelle2}
\begin{equation}
I=\int d^{4}x\sqrt{-g}\left\{ \frac{1}{ 2 \kappa_0}\left(R-2\Lambda_{0}\right) + \alpha R^2 + \beta R_{\mu \nu} R^{\mu \nu} \right\},
\label{action2}
\end{equation}
one realizes that the maximally symmetric vacuum is not unique (there are now two vacua in general) and that in addition to the massless graviton, there is a massive spin-two ghost and a massive spin-0 graviton. All these modes contribute to gravitation in the weak field limit which could be attractive or repulsive depending on the parameters.\footnote{ One can design theories such as critical gravity with tuned parameters that are free of tree-level ghosts, but these are rather special theories relying on the existence of a tuned cosmological constant \cite{pope_critical,tahsin_critical}.}  One can study (\ref{action2}) as an interim theory "outcasting" the problem of the presence of the ghost to the higher orders and to the microscopic quantum gravity at the end (see the recent surge of activity about solutions of this theory \cite{Nelson,Pope1,Pope2,Lust}.)  While this is a perfectly legitimate point of view, it simply 
makes the predictions of the theory unreliable at the scales where  higher derivative terms make important contributions. More importantly, since the  quadratic terms make contributions to the propagator, even the tree-level result namely the particle spectrum, is questionable.   Therefore, in a series of papers  \cite{tahsin1,tahsin2,tahsin3}, an alternative, a bottom up, approach to building low energy quantum gravity actions were suggested. Namely, in these works  the following question was answered:
\begin{itemize}
\item  Can one built extended theories with arbitrary powers of curvature that have a unique maximally symmetric vacuum and a single massless spin-2 excitation, just like Einstein's gravity, about the vacuum ?
\end{itemize}
Such theories are easily found if the vacuum is flat but these theories are not that restricted that practically there are as many arbitrary parameters as there are powers of curvature. Hence  a stronger condition is needed: that is the uniqueness of the (A)dS vacuum and the existence of a unitary massless graviton about this vacuum.  These two conditions are quite hard to satisfy but still one needs a further requirement to restrict the set of theories. This requirement is the condition of minimality, that is, no derivatives of the Riemann tensor and its contractions should appear \cite{Gibbons}. This idea is borrowed from pre-quantum electrodynamics era quantum field theories that appear in the form of Born-Infeld or Euler-Heisenberg actions. As  the details of the construction and motivation are laid out in the works \cite{tahsin1,tahsin2,tahsin3,Tahsin_tez}, we simply quote the final theory obtained in a tour de force that minimally extends Cosmological Einstein's gravity in four dimensions.\footnote{We rescale $\gamma  \rightarrow 4 \gamma$ compared to the one in \cite{tahsin2,tahsin3}. The theory takes a particularly elegant form in 2+1 dimensions \cite{bi3}.}
\begin{equation}
I=\frac{1}{2 \kappa_0\gamma}\int d^{4}x\,\left[\sqrt{-\det\left(g_{\mu\nu}+4\gamma A_{\mu\nu}\right)}-\left(4 \gamma\Lambda_{0}+1\right)\sqrt{-\det g}\right],\label{eq:Generic_BI_intro}
\end{equation}
where the minimal two tensor that satisfies the constraints is 
\begin{eqnarray}
&A_{\mu\nu}=  R_{\mu\nu} + c S_{\mu\nu} \nonumber \\
& + 4 \gamma \left( a C_{\mu\rho\nu\sigma} R^{\rho\sigma} + \frac{c + 1}{4} R_{\mu\rho} R_{\nu}^{\rho} + \left( \frac{c \left(c+2\right)}{2} - 2 - b \right) S_{\mu\rho} S_{\nu}^{\rho} \right) \nonumber \\
& +\gamma g_{\mu\nu} \left( \frac{9}{8} C_{\rho\sigma\lambda\gamma} C^{\rho\sigma\lambda\gamma} - \frac{c}{4} R_{\rho\sigma} R^{\rho\sigma} + b S_{\rho\sigma} S^{\rho\sigma} \right), 
\label{EGB_Amn} 
\end{eqnarray} 
where $a$, $b$ and $c$ are arbitrary real dimensionless parameters and $\gamma$ is the Born-Infeld parameter with dimensions of $[L^2]$.   Here $C_{\mu \nu \sigma \rho}$ is the Weyl-tensor and $S_{\mu \nu}$ is the traceless Ricci tensor. For any values of the 3 dimensionless parameters and $\gamma >0$, the theory has the following properties ( with $\lambda \equiv \gamma \Lambda$ and $\lambda_0 = \gamma \Lambda_0$ )

\begin{itemize}

\item With a given $\lambda_{0}<\frac{11}{64}$, it has a unique viable
maximally symmetric vacuum with a cosmological parameter $\lambda< \frac{1}{4}$,
and an effective Newton's constant
\begin{equation}
 \frac{1}{{\kappa}}=\frac{1}{{\kappa}_0}\left(1-4 \lambda\right)\left(1+2\lambda\right)^{2}.
\label{eff_newton}
\end{equation}
\item It describes a unitary massless spin-2 excitation around this vacuum
for any value of $\lambda_{0}<\frac{11}{64}$ including $\lambda_{0}=0$,
except $\lambda_{0}=-1/4$, which yields $\lambda=-1/2$ and so ruled
out by the requirement of a non-zero effective Newton's constant.

\item The effective cosmological parameter is determined from the equation
\begin{equation}
4 \lambda^4 + 4 \lambda^3 - \lambda + \lambda_0=0,
\label{eff_cosmo}
\end{equation}
which a priori has  four roots but it turns out that out of the two real roots, only one of them is viable in the sense that the graviton is unitary around this vacuum and non-unitary (a ghost) around the other one.  
So the theory has the same perturbative spectrum (propagator) as the cosmological Einstein theory.

\item The square-root nature of the  Lagrangian is important: That is, when expanded in curvature, it provides an infinite order unitary extension of Einstein's gravity.  What is rather remarkable, is that beyond the fourth order terms in the curvature, no new constraints and contributions appear in the effective parameters and the 
vacuum equation. Namely, effective Newton's constant is given as (\ref{eff_newton}) and the effective cosmological constant is determined from (\ref{eff_cosmo}). Hence, one can study any truncated version of the action (\ref{eq:Generic_BI_intro}) as a unitary  action with a unique viable vacuum and a massless spin-2 graviton whose free Lagrangian properties are given as Einstein's gravity with an effective Newton's constant and an effective cosmological constant. 

\item If one just wants to go up to and including order  $ {\cal O}(R^3)$ truncation, then one still has a theory with a unique maximally symmetric vacuum and a unitary massless graviton as long as   $ -\frac{1}{3 \sqrt{3}} < \lambda_0 <  \frac{11}{64}$ with  $-\frac{1}{2 \sqrt{3}} < \lambda <\frac{1}{4}$. Hence a lower bound is introduced to these parameters. 

\item At ${\cal O}(R^2)$, the theory reduces to the Einstein Gauss-Bonnet theory. Since the Gauss-Bonnet term is a total derivative in four dimensions, field equations are just the same as Einstein's gravity.  But, the Gauss-Bonnet term plays a major role in the construction of the actions of the Born-Infeld form.

\end{itemize}

In what follows, we shall study the spherically symmetric solutions  and some general futures of a restricted version of the above theory (\ref{eq:Generic_BI_intro}) to see in more explicit terms how Einstein's gravity is improved. This restricted version has no free arbitrary parameters. they are fixed as $c= -1$, $a =0$ and $b = - \frac{5}{2}$. This particular choice of the dimensionless parameters is quite interesting since, only for this  choice, the theory reduces to the  following form free of the square root: 
\begin{equation}
I= \frac{1}{ 2\kappa_0} \int d^{4} x \sqrt{-g}\, {\cal F}( R, {\cal{G}} ) ,
{\label{fourthorder}}
\end{equation}
where the Lagrangian density reads 
\begin{equation}
{2 \gamma} {\cal F}  \equiv \left(1 + \gamma R -\frac{1}{2} \gamma^2 \left(R^2- 9\,  {\cal{G}} \right) \right)^2-4 \lambda_0 -1. 
\label{lag_f}
\end{equation}
In passing from (\ref{EGB_Amn}) to (\ref{lag_f}), we have made use of the identity in four dimensions relating the Gauss-Bonnet combination
\begin{equation}
{\cal {G} } \equiv   R_{\mu \nu \sigma \rho} R^{\mu \nu \sigma \rho} - 4 R_{\mu \nu}   R^{\mu \nu} + R^2, \end{equation}
to the square of the Weyl-tensor  via
\begin{equation}
C_{\mu \nu \sigma \rho} C^{\mu \nu \sigma \rho} = {\cal {G} } +2 R_{\mu \nu}   R^{\mu \nu} - \frac{2}{3} R^2 . 
\end{equation}
In the light of the above discussion, the fourth order theory (\ref{lag_f}) is a minimal extension of Einstein's gravity having a unique viable vacuum and a massless spin-2 excitation about it, with no other modes. Moreover, this statement is true if ${\cal O}(R^4)$ terms are simply  dropped from the action and only ${\cal O}(R^3)$ truncation is kept. As these, not so obvious, conclusions were derived with somewhat infrequently used techniques (such as building equivalent linear and quadratic actions that have the same properties ) in the works \cite{tahsin1,tahsin2,tahsin3}, in what follows,  before we move on to the discussion of solutions of this theory, we shall find the vacuum and the particle spectrum of the full quartic theory and its cubic truncation with the more conventional techniques of finding the field equations, searching for the maximally symmetric solutions and linearizing about the vacua to identify the particle spectrum. It will be refreshing to see that the hard to satisfy conditions about the uniqueness of the vacuum and the unitarity of the massless spin-2 graviton with no other massless or massive modes are fulfilled in this theory.

\section{ Vacuum and Spectrum of the Theory}

Bluntly varying the action (\ref{fourthorder}) with respect to the metric leads to highly complicated field equations as it is a quartic theory in the curvature. Hence it is a good idea to make use of partial derivatives of the Lagrangian with respect to two scalar  curvature invariants: the scalar curvature $R$ and the Gauss-Bonnet combination ${\cal{G}}$. This procedure is also somewhat long but it eventually yields the following rather compact equations 
     \begin{align}\nonumber
        & {\cal F}_R  R_{\mu \nu}+\frac{1}{2}g_{\mu \nu}
        ({\cal {G}} {\cal F}_{\cal G}- {\cal F}) +\Big (g_{\mu \nu}\square -\nabla_\mu \nabla_\nu \Big )  {\cal F}_R \\
        & +4\left[ \Big (2 C_{\mu \sigma \nu \lambda}-R_{\mu \sigma \nu \lambda} \Big )\nabla^\sigma \nabla^\lambda
        +\frac{R}{6}\Big (g_{\mu \nu}\square -\nabla_\mu \nabla_\nu \Big )\right] {\cal F}_{\cal G} =  8 \pi G_0 T_{\mu \nu},
\label{full_eqn}     
    \end{align}  
where  the two relevant partial derivatives are defined and computed for our theory as 
\begin{equation}
 {\cal F}_{\cal G} \equiv \frac{\partial  {\cal F}}{\partial {\cal G}} = \frac{9}{4}  \gamma \left( -\gamma^2 R^2+9 \gamma^2 {\cal G} +2 \gamma R +2\right),
\end{equation}
\begin{equation}
 {\cal F}_R \equiv \frac{\partial  {\cal F}}{\partial R} =\frac{1}{2} \Big (\gamma R-1 \Big ) \Bigg (\gamma R \Big (\gamma R-2 \Big)-9 \gamma^2 {\cal{G}} -2 \Bigg ).
\end{equation}
We have also added a source term and set $\kappa_0 \equiv 8 \pi G_0$. Note that $G_0$ is the bare Newton's constant which is not the one measured in the lab. The one measured in the lab receives corrections from the cosmological background and is given by (\ref{eff_newton}) and only reduces to $G_0$ in flat backgrounds. 

The trace of the field equations will also be used which is 
    \begin{equation}
     R {\cal F}_R+2{\cal {G}} {\cal F}_{\cal {G}}- 2{\cal F}+3 \square {\cal F}_R-4 G_{\mu \nu}\nabla^\mu \nabla^\nu {\cal F}_{\cal {G}} 
       =  8 \pi G_0 T.
\label{trace_field}     
 \end{equation} 
 Let us now find the maximally symmetric solutions.  A priori there could be at most four of them as it is a quartic theory. Setting $T_{\mu \nu}=0$ and denoting the metric and the related tensors with an over-bar for the maximally symmetric solution, one has 
\begin{equation}
          \bar{R}_{\mu \sigma \nu \rho}=\frac{\Lambda}{3}(\bar{g}_{\mu \nu}\bar{g}_{ \sigma \rho}-\bar{g}_{\mu \rho}\bar{g}_{\sigma \nu}),
           \end{equation}
where $\Lambda$ is the effective cosmological constant to be determined in a moment.
  The trace equation (\ref{trace_field}) reduces to 
    \begin{equation}
              \bar{R}\,\bar{\cal F}_R+2 \bar{{\cal {G}}}\,\bar{\cal F}_{\cal {G}} -2\bar{\cal F}=0,
\label{bartrace}               
\end{equation} 
where one has  $\bar{R} = 4 \Lambda$, $\bar{ \cal G} = \frac{8}{3} \Lambda^2$ and  the background values  of the involved functions read 
\begin{equation}
     \bar{\cal F}_R = (1- 4 \lambda) \Big(1+2 \lambda \Big)^2, \hskip 0.5 cm     \bar{\cal F}_{\cal {G}}= \frac{9}{2}\gamma  ( 1 + 2 \lambda )^2, \hskip 0.5 cm  \bar{\cal F}=\frac{1}{2 \gamma} \left[( 1 + 2\lambda)^4 -4\lambda_0-1\right],
\end{equation}
where we used the dimensionless cosmological parameter $\lambda$ wherever it is possible. 
Inserting these in (\ref{bartrace}) one arrives at the desired equation
\begin{equation}
4 \lambda^4 + 4 \lambda^3 - \lambda + \lambda_0=0,
\label{vac_denk}
\end{equation}
which is of course solvable but the general solution is neither particularly illuminating to look at and nor it is necessary. For our purposes, the more important issue is the uniqueness of a viable solution. Namely, we do not want more than one good solution.  Therefore, we must study the constraints on the involved parameters. 
To this end, the "discriminant" is a useful tool which reads
\begin{equation}
\Delta = 16 ( 1 + 4 \lambda_0)^2 ( -11 + 64 \lambda_0).
\label{disc}
\end{equation} 
At this stage uniqueness of the viable  vacuum is not yet established: Common knowledge on the solutions of quartic equations show that depending on the value of $\lambda_0$, there could be more than one real solution with the discriminant given as above. To explore the possibilities and the conditions on the parameters, we need to study the excitations about all potential vacua  of the theory. So, say generically $\lambda$ is a viable effective cosmological constant of the theory.  Then let us consider, perturbations, (later to be identified as spin-2 excitations ) about it as 
 \begin{equation}
   g_{\mu \nu}= \bar{g}_{\mu \nu}+h_{\mu \nu}.  
\end{equation}             
  Then  using the subscript (or superscript) $L$ to refer to linearized quantities about the vacuum, the linearization of the Gauss-Bonnet invariant is given as {\cite{Deser_Tekin}
  \begin{equation}
       {\cal {G}}_L  =\frac{4}{3}\Lambda R_L,
\end{equation} 
where $R_L$ is the linearized part of the scalar curvature, which more explicitly, reads 
 $ R_L = - (\bar{\square} + \Lambda) h + 
\bar{\nabla}^\mu \bar{\nabla}^\nu h _{\mu \nu}$, with  $h \equiv \bar{g}^{\mu \nu} h_{\mu \nu}$.
One can also compute the following linearized quantities for the quartic gravity
 \begin{equation}
 ( {\cal F}_R)^L  = -6  \gamma \lambda \Big(1+2 \lambda \Big)R_L, \hskip 0.5 cm  
( {\cal F}_{\cal G})^L  = \frac{9}{2} \gamma^2 \Big(1+2 \lambda \Big)R_L\hskip 0.5 cm ( {\cal F})^L  =  \Big(1+2 \lambda \Big)^3 R_L . 
\label{linearized_quant}    
 \end{equation}
It is a good idea to first look at the  linearization of the trace of the field equations which potentially hides a possible spin-0 mode that we do not want in our theory as we want to keep the spectrum of Einstein' s gravity intact. The linearization of the trace equation reads
\begin{align}\nonumber
 &({\cal F}_R)^L \bar{R}+\bar{\cal F}_R R_L+2{\cal {G}} _L\bar{\cal F}_{\cal {G}}+2\bar{{\cal {G}}} ({\cal F}_{\cal {G}})^L-2{\cal  F}_L \\&+3\bar{\square}({\cal F}_R)^L
       -4\bar{R}_{\mu \nu}\bar{\nabla}^\mu\bar{\nabla}^\nu ({\cal F}_{\cal {G}})^L+2\bar{R} \,  \bar{\square}({\cal F}_{\cal {G}})^L=0.
\end{align} 
Upon use of (\ref{linearized_quant}), one arrives at the crucial point here :The second line in this equation vanishes identically. Hence the $\bar{\square} R_L$ term drops out and the linearized part of the trace equation does not give rise to a wave-like equation. This means that there will not be a spin-0 mode in this theory, as mentioned before. The remaining part of the trace equation yields 
\begin{equation}
( 1+ 2 \lambda)^2( -1 + 4 \lambda)R_L=0.
\label{tr_lin}
\end{equation}
Since $R_L$ is a background diffemorphism invariant  ( "gauge-invariant" )  object under the transformations $\delta_\zeta h_{\mu \nu} = \bar{\nabla}_\mu \zeta_\nu+ \bar{\nabla}_\nu \zeta_\mu$, it cannot be left undetermined in this theory. Therefore, from  (\ref{tr_lin}) one concludes that $ \lambda \ne \frac{1}{4}$, $ \lambda \ne - \frac{1}{2}$ and $R_L =0$. Let us move ahead and linearize the full equation (\ref{full_eqn}) in the absence of a source with the condition $R_L=0$, which simply yields
\begin{equation}
(1- 4 \lambda) (1 + 2 \lambda)^2 \Big ( R_{\mu \nu}^L - \Lambda h_{\mu \nu} \Big ) = 0,
\label{effN}
\end{equation}
which also shows clearly the correctness of the exclusion of two specific values of $\lambda$ as notes above. Note that since $R_L=0$, the linearized form of the cosmological Einstein tensor is $G_{\mu \nu}^L =   R_{\mu \nu}^L - \Lambda h_{\mu \nu}$. Therefore, our quartic theory  simply reduces to the cosmological Einstein's theory at the linearized level, with a single massless spin-2 excitation.  To be able to read 
the effective Newtons constant, we can couple the linearized equation to a weak source which yields 
the effective Newton's constant as 
\begin{equation} 
\frac{1}{ G_{\mbox{eff}} }= \frac{1}{G_0}  (1- 4 \lambda) (1 + 2 \lambda)^2 .
\end{equation}
So clearly, given $G_0 >0$, which is required for attractive gravity in the flat space limit of the theory, we must have  $\lambda < \frac{1}{4}$ and $\lambda \ne -\frac{1}{2}$.  Choosing the transverse traceless gauge  ( $h = 0$, and  $\bar{\nabla}_\mu h^{\mu \nu} =0$ ) which is compatible with the condition $R_L=0$, one arrives at the massless spin-2 equation in (A)dS in the absence of sources $ \Big (\bar{\square} - \frac{2}{3}\Lambda \Big)h_{\mu \nu}=0$. 

The knowledge we gained from the analysis of the perturbative spectrum about the putative  maximally symmetric  solution is sufficient to go back and identify the unique viable solution of (\ref{vac_denk}).  To determine the allowed  range of $\lambda_0$, let us rewrite that equation as 
\begin{equation}
\lambda_0= \lambda- 4 \lambda^3  - 4 \lambda^4 , 
\label{lambda0}
\end{equation}
 with the first and second derivatives given as
\begin{equation}
\frac{d \lambda_0}{d \lambda} = ( 1 + 2 \lambda)^2( 1- 4 \lambda), \hskip 1 cm \frac{ d^2 \lambda_0}{d \lambda^2} =-24 \lambda ( 1 + 2 \lambda).
\end{equation}         
Since, $\lambda \ne -\frac{1}{2}$, there is a single maximum at $\lambda = \frac{1}{4}$, which is not also physically allowed as we have seen above, but this value when used in (\ref{lambda0}) leads to the condition $\lambda_0 < \frac{11}{64}$. Using this in the discriminant (\ref{disc}), we get $\Delta <0$ and the well-known analysis of the quartic equation leads to the conclusion that there are two real and two complex conjugate roots. It is easy to show that one of the real roots is in the allowed region $\lambda <1/4$, while the other one is always out of the allowed region $\lambda >1/4$. The theory has the remarkable property 
that it has a unique vacuum and a single massless unitary spin-2 graviton about this vacuum. As we have discussed the virtues of the unique vacuum in gravity elsewhere \cite{tahsin2,tahsin3}, we do not repeat that  discussion here but simply note that unless a natural selection mechanism is found (which does not seem to exist currently) between two different maximally symmetric vacua in gravity, one best has a theory with a unique vacuum, hence part of the motivation of this and related work we have done.

At this stage we can summarize the conditions coming from the existence of a maximally symmetric solution and the unitarity  (non-ghost nature) of the massless spin-2 particle  as  $ \lambda_0 < \frac{11}{64}$ and $\lambda < \frac{1}{4}$ with $\lambda \ne -\frac{1}{2}$. So, once $\lambda_0$ is given, one can solve for $\lambda$.  We can write the full solution for $\lambda$ but it is cumbersome and not really needed. Instead, we plot Figure 1  where the vertical axis is $\lambda$ and the horizontal axis is $\lambda_0$.  At this stage there are no lower bounds on the values of the parameters. But one thing to notice is the following, taking a very large negative value of $\lambda_0$, since $\lambda \approx ( -\lambda_0)^{1/4}$, the effective cosmological constant is significantly smaller than the bare one.  
\begin{figure}[h]
\centering
\includegraphics[width=0.5\textwidth]{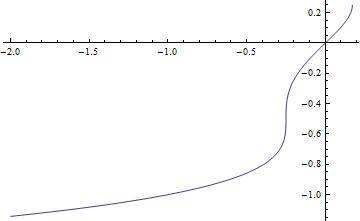}
\caption{Depiction of $\lambda$ vs $\lambda_0$  in the allowed region. Note that at $\lambda_0=-1$, $\lambda=-1$}
\label{hyp_embed}
\end{figure}

In Figure 2, we plot $G_{\mbox{eff}}/G_0$ versus $\lambda$.  The points $\lambda =-0.5$ and $\lambda =0.25$ are excluded as we have seen above. It is interesting to see that as $\lambda$ gets  larger in magnitude, the effective Newton's constant gets smaller. Basically one has $G_{\mbox{eff}} \approx  \frac{G_0}{8 |\lambda|^3 } \approx \frac{G_0}{8 |\lambda_0|^{\frac {3}{4} }}  $ .
\begin{figure}[h]
\centering
\includegraphics[width=0.5\textwidth]{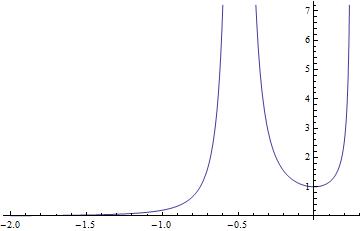}
\caption{Depiction of $G_{\mbox{eff}}/G_0$ versus $\lambda$ for $ -2 < \lambda < 0.25$. Note that $\lambda = -0.5$ is not allowed.}
\label{hyp_embed}
\end{figure}

One further remark is apt here: We can demand more from our theory; by simply dropping the  ${\cal O}(R^4)$ terms from the action (\ref{lag_f}), we arrive at a cubic truncation 
\begin{equation}
{\cal F}= R -2 \Lambda_0 +\frac{9 \gamma  {\cal G} }{2}+\frac{9}{2} \gamma ^2  {\cal G}
   R-\frac{1}{2} \gamma ^2 R^3.
\end{equation} 
As usual, the third term can be dropped from the action as it does not contribute to the field equations, but we shall keep it. We can run the same machinery for this theory as we did for the full quartic theory and get the spectrum and the vacuum.  It is easy to see that the vacuum equation becomes 
\begin{equation}
 4 \lambda^3 - \lambda + \lambda_0=0,
\label{vac_denk3}
\end{equation}
and the unitarity condition of the massless spin-2 excitation becomes
\begin{equation}
G_{\mbox{eff}} = \frac{G_0}{ 1 -12 \lambda^2} >0,
\end{equation}       
which restricts $\lambda$ to the interval   $ - \frac{1}{2 \sqrt{3}} < \lambda < \frac{1}{2 \sqrt{3}}$.  Since the upper bound is weaker than the bound coming from the quartic theory ( which was $\lambda < \frac{1}{4}$) it is already satisfied, but a lower bound on the effective cosmological parameter appears.  Let us now look at the allowed range of $\lambda_0$. For the allowed region of $\lambda$, it is easy to see that one has 
$ -\frac{1}{3 \sqrt{3}}< \lambda_0 < \frac{3}{16}$. Since the upper limit is weaker that  the earlier limit of $\frac{11}{16}$, it does not bring any new constraint. But of course, the lower bound is new.  All that remains is that we have to show that there is a solution to (\ref{vac_denk3}) in this region and hopefully this solution is unique. Computing the relevant discriminant we get $\Delta =  16 ( 1- 27 \lambda_0^2)$, which says that in the allowed region $\Delta <0$ and there is a single real solution to the cubic equation.  

With the same reasoning we require our theory to be unitary at ${\cal O}(R^2)$ and at  ${\cal O}(R)$ but the former is just the Einstein-Gauss-Bonnet theory with a cosmological constant  and the latter is the Einstein's theory with a cosmological constant. No new conditions arise from the lower order truncations besides the positivity of the bare Newton's constant $G_0>0$ which we already used. Moreover the vacuum is unique with $\lambda = \lambda_0$.  Hence the basic premise of the works  \cite{tahsin2,tahsin3} of building unitary theories at order  $ {\cal O}(R^n)$ and at  every truncated power $ m < n$ is fulfilled in this setting of $n= 4$.  We would like to stress  the requirement that the theory at hand be tree-level unitary with only a massless spin-2 excitation about its unique vacuum at every order in all possible effective truncations is a rather novel concept introduced in  \cite{tahsin2,tahsin3}.  In the literature, one will often find that truncated versions of a theory has problems such as ghosts etc. which are expected to disappear in the full microscopic theory.  This is not the point of view pursued in our work. For arbitrary $n$, we refer the reader to these two quoted works as our next task here is to carry out a detailed study of the spherically symmetric solutions of the quartic theory.

\section{Einsteinian solutions}

Let us now consider the Ricci flat metrics, namely the Einsteinian solutions with  $R_{\mu \nu}=0$ for  the case of $T_{\mu \nu} =0$, $\lambda_0=0$. This is an important discussion as we would like to know which solutions of Einstein's gravity are inherited in our quartic gravity theory. For Ricci flat metrics,  we have 
\begin{equation}
 {\cal F} =  \frac{9 \gamma  {\cal G}}{2} \left(1  + \frac{9}{4} \gamma^2 {\cal G} \right), \hskip 0.5 cm 
 {\cal F}_{\cal G} =\frac{9 \gamma }{2} \left(1  + \frac{9}{2} \gamma^2 {\cal G} \right),\hskip 0.5 cm
 {\cal F}_{R} =1  + \frac{9}{2} \gamma^2 {\cal G}.
\end{equation}
Here the Gauss-Bonnet invariant simply reduces to the Krestchmann scalar  ${\cal G}= R_{\mu \nu \alpha \beta}R^{\mu \nu \alpha \beta}$. The trace of the field equations (\ref{trace_field}) reduces to a non-linear wave equation as
\begin{equation}
\square{\cal G}+ \frac{3}{2}\gamma {\cal G}^2=0.
\label{non_lin}
\end{equation}
This is already a rather impressive result: in Einstein's theory,  Kretschmann scalar  is computed once the solution is found. Namely,  there is no separate equation satisfied by this curvature invariant. For example, for the Schwarzschild solution or the Kerr solution (which are Ricci flat),  ${\cal G}$ is singular and one must live with this fact within the framework of the classical theory. But here, this curvature invariant satisfies an equation on its own and hence singular solutions, if there are any, need not be  accepted on the basis of physics.  Of course, this does not say that there are no other curvature invariants which could be singular. But in any case, the fact that the Kretschmann scalar satisfies a non-linear wave equation on its own for all Ricci flat solutions is highly non-trivial as rotating and non-rotating black hole solutions or in fact any solution of Einstein's theory in vacuum without a cosmological constant falls into this category. [Note that when there is a bare cosmological constant, the above equation simply picks up an inhomogeneous term as $ \square{\cal G}+ \frac{3}{2}\gamma {\cal G}^2 = 4 \Lambda_0$ without much change of the conclusions in this paragraph.]  Finally, let us also write the other part, namely the traceless part, of the field equations for Ricci flat metrics:
\begin{equation}
 \Bigg ( \frac{1}{4} g_{\mu \nu} \Box  - \nabla_\mu \nabla_\nu \Bigg ) {\cal G} + 18 \gamma C_{\mu \sigma \nu \rho} \nabla^\sigma \nabla^\rho {\cal G} =0.
\label{tr_less}
\end{equation}
Therefore for a Ricci flat metric to be a solution to the quartic gravity,   ${\cal G}$ must satisfy this equation and the non-linear Klein-Gordon equation (\ref{non_lin}). One obvious solution is the flat space itself. 
Which solutions of Einstein's gravity survive as solutions to this theory is an outstanding problem itself. It is very likely that only very few solutions without singularities will survive. Among these are the exact  waves which solve generic gravity theories {\cite{Gurses_prl, Gurses_prd}.

Searching for solutions to the non-linear wave equation (\ref{non_lin}) that must also solve (\ref{tr_less})  is a non-trivial problem on its own and is beyond the scope of this work, which we shall address in a separate work.  But let us show that there could be static black hole solutions (by a black hole, we mean that there is an event horizon) along the discussions of  {\cite{Nelson,Pope1,Pope2}. 
We can take the most general static metric as 
\begin{equation}
 ds^2=-N^2 dt^2+h_{ab}dx^a dx^b,
 \end{equation}
with $N=N(x^a)$ and $h_{ab}=h_{ab}(x^a)$ depending on the spatial coordinates only, yielding 
the components of the Christoffel connection as   $\Gamma ^0 _{0i}=\partial _i \ln N$ and  $\Gamma ^i _{00}= N h^{i j} \partial _j N$. So given a scalar function $\psi(x^a)$, one can split the wave operator as 
\begin{eqnarray}\nonumber
\square \psi (x_a) &=& g^{\mu \nu} \nabla_\mu \nabla_\nu \psi
= g^{00} \nabla_0 \nabla_0 \psi+h^{ab} \nabla_a \nabla_b \psi\\\nonumber
&=& - g^{00} \Gamma ^i _{00}\nabla_i \psi +h^{ab} \nabla_a \nabla_b \psi\\\nonumber
&=& h^{ab} \nabla_a \nabla_b \psi+\frac{1}{N}h^{ab}\partial_b N \nabla_a \psi.
\end{eqnarray}
Denoting the 3 dimensional covariant derivative (which is compatible with the spatial metric $h_{a b}$ ) by $D_a$, and carrying out the lowering and raising with the spatial metric, we arrive at 
\begin{equation}
\square \psi (x_a) =  D^a D_a \psi+\frac{1}{N}(D^a{N}) D_a  \psi.
\end{equation}
Applying this to the non-linear Klein-Gordon equation, one finds
\begin{equation}
D^a D_a {\cal G}+ \frac{1}{N}D^a N D_a {\cal G}+\frac{3}{2}N{\cal G}^2=0, 
\end{equation}
which is valid for all static Ricci flat solutions. Multiplying this by $N{\cal G}$ and integrating over the 3-dimensional spatial section, one gets
\begin{equation}
\int _S \sqrt{h}d^3 x \left[ N{\cal G}D^a D_a {\cal G}+{\cal G}
D^a N D_a {\cal G}+\frac{3}{2}\gamma N {\cal G}^3\right] =0, 
\end{equation}
which upon organization of the terms reduces to 
\begin{equation}
\int _S \sqrt{h}d^3 x \left[D^a \left( N{\cal G}D_a {\cal G}\right) -
N D^a {\cal G}D_a {\cal G}+\frac{3}{2}\gamma N{\cal G}^3\right] =0. 
\end{equation}
Consider now $S$ to be the spatial region  between the event horizon of the assumed black hole and  the spatial infinity. Then, by definition, $N$ vanishes on the horizon and the integral of the first term yields zero on both asymptotics, for asymptotically flat spaces as ${\cal G} \rightarrow  {\cal O}(1/r^{2+ \epsilon})$. But the remaining integrand is not positive definite or negative definite, hence  ${\cal G}$ need not vanish in the bulk, outside the horizon, and hence static black hole solutions are not ruled out.  

To avoid a possible misunderstanding, let us note that the discussion in this section does not in any way prove that the quartic theory presented here is a singularity-free theory in the sense that a given initial matter distribution satisfying some kind of reasonable energy condition will not yield a geodesically-incomplete spacetime as laid out by Hawking and Penrose \cite{Hawking} for general relativity.  
This question is still open, the theory we have is a higher order one which might be less singular or even be non-singular but  showing that requires another work which is beyond the scope of the current attempt where we have only shown that, as the title of this section shows, the singular solutions to the vacuum Einstein equations ($R_{\mu \nu} =0$), have to satisfy a wave-like equation whose singular solutions can be eliminated on account of physical grounds.

Let us now turn our attention to approximate spherically symmetric static  solutions which surely exist almost by construction since at large distances, Einstein's gravity is dominant. 

\section{Spherically Symmetric Solutions }

Let us consider the spherically symmetric ansatz
\begin{equation}
ds^2  = - g(r)^2 f(r) dt^2 + \frac{1}{f(r)} dr^2 + r^2 ( d \theta^2 + \sin^2 \theta d \varphi^2 ).
\end{equation}
Inserting this ansatz into the action (\ref{fourthorder}), and integrating out the  inconsequential (for the purpose of calculus of variations) angular and the time parts  yields the following result 
\begin{equation}
I_{\mbox{reduced}} = \int_0^\infty  x^2 dx  g(x) \Big (  A(x)^2   -1 \Big),
\label{reduced_action}
\end{equation}
where we have defined a dimensionless variable  $x \equiv \frac{r}{2 \sqrt{\gamma}}$ and  $A(x)$ is computed to be  
\begin{eqnarray}\nonumber
A(x) &=& 1-\frac{1}{8x^4}\Big(1-4x^2-2f +4x^2 f \Big)
-\frac{ f^2}{8 x^4}\Big(1+4x^2g'^2 +x^4 g''^2\Big) \\\nonumber
&-&  f'\Big(\frac{1}{x}-\frac{1}{2x^3}+\frac{f}{2x^3}-\frac{5f'}{8x^2} \Big)      
-   \frac{f^2 g' }{2x^3} \Big( g+x^2 g''     \Big)    \\\nonumber
&-&    f'g' \Big( \frac{3g}{4} + \frac{3g}{x^2}+ \frac{3gf'}{4x} +\frac{9f'g'}{32}+\frac{3f''g}{16}\Big)  
+ \frac{ff'g'}{8x^2}\Big(  34g-6xg'-3x^2g''\Big) \\\nonumber
&-& \frac{fgg'}{4x}\Big(4-\frac{2}{x^2}+f''\Big)  
-\frac{f''}{4x^2}  \Big( 4+x^2-4f+xf'+\frac{x^2 f''}{8}  \Big)  \\
&-&\frac{2fgg''}{x^2}\Big(1+\frac{ x^2}{4}-f+\frac{ xf'}{4}+\frac{x^2 f''}{16}\Big).
\end{eqnarray}
We have also set the bare cosmological constant to zero. According to the rigorously proven notion of "symmetric criticality", it is well-known \cite{Palais, shortcuts} that, succinctly speaking   "critical symmetric points are symmetric critical points" when the integrated symmetry group is compact, such as the one here. Therefore no equation is lost when the above action is varied with respect to the two metric functions $f(x)$ and $g(x)$.  Note that with the help of a computer program, such as Mathematica, this method if often faster in these higher derivative theories. This shortcut leads to two complicated equations which we do not depict here. But, we shall note our main results regarding the solutions of these coupled non-linear equations.

\begin{itemize}

\item  Minkowski space is a solution. Namely  $f(r) = 1$ and $g(r) =1$. 

\item The Schwarschild metric is not an exact solution. There are various ways to see this result. It follows from the two equations coming from the variation of the reduced action (\ref{reduced_action}), but, as it is rather illuminating let us give another derivation, directly coming from the full set of equations, here. Denoting  $ {\cal E}_{\mu \nu} =0$ as the vacuum equations of the theory, one observes that when the Schwarzschild metric  ( $g (r) =1$ and $f(r) = 1- \frac{ 2 G m}{r}$ ) is inserted one finds that it fails to be a solution as follows
\begin{eqnarray}
 &&{\cal E}_{tt} =-\frac{1296 \gamma ^2 G^2 m^2 (2 G m-r) \left(r^3 (11 G m-5 r)+9 \gamma  G m (67 G
   m-32 r)\right)}{r^{13}} \approx {\cal O}( \frac{1}{r^8}) \nonumber \\
&&{\cal E}_{rr} =\frac{1296 \gamma ^2 G^2 m^2 \left(r^3 (2 r-3 G m)+9 \gamma  G m (11 G m-4
   r)\right)}{r^{11} (2 G m-r)} \approx {\cal O}( \frac{1}{r^8}) \nonumber \\
 &&{\cal E}_{\theta \theta}=\frac{1296 \gamma ^2 G^2 m^2 \left(2 r^3 (3 r-7 G m)+9 \gamma  G m (41 G m-18 r)\right)}{r^{10}} \approx {\cal O}( \frac{1}{r^6})
\end{eqnarray}
and  $  {\cal E}_{\varphi \varphi}  = {\cal E}_{\theta \theta} \sin^2 \theta$. This failure of the Scwarzshild metric  at ${\cal O}(\frac{1}{r^6})$ to be a solution also provides one a way to construct approximate solutions at large distances which can be easliy found as 
\begin{equation}
g(r) = 1+ {\cal O}( 1/r^8),\hskip 0.5 cm f(r) =1-\frac{2 G m}{r}-\frac{2592 G^2 m^2 \gamma^2 }{5 r^6}+ \frac{864 G^3 m^3 \gamma^2 }{r^7} + {\cal O}( 1/r^8).
\end{equation}  
Notice that  at  and beyond $  {\cal O}( 1/r^8)$, one does not have  $g_{00} g^{rr} = -1$ anymore as the metric function $g(r)$ differs from 1.   To go beyond  $  {\cal O}( 1/r^8)$, one must carefully reorganize the power series expansion as the theory is a non-linear one. From the full equations, one finds that the solution up to  $  {\cal O}( 1/r^9)$, reads 
\begin{eqnarray}
&&f(r)= 1-\frac{2 G
   m}{r} +\frac{864 \gamma ^2 G^3 m^3}{r^7}+\frac{1296 \gamma ^2 G^2 m^2}{r^6}-\frac{486
   \left(1615 \gamma ^2 G^4 m^4-864 \gamma ^3 G^2 m^2\right)}{133 r^8} \nonumber \\
&&g(r)=1 -\frac{756 \gamma ^2 G^2 m^2}{r^6}  -\frac{9720 \gamma ^2 G^3 m^3}{7 r^7}-\frac{26244 \gamma ^3 G^2 m^2}{19
   r^8}.
\end{eqnarray}
\end{itemize}
These constitute the approximate  solution to the Schwarschild metric outside the event horizon.

\section{Conclusions}

In this work we have pursued further the recently {\cite{tahsin2,tahsin3} introduced  idea that "the purely gravitational sector of low energy quantum gravity theories should have the same properties as Einstein's theory, as far as their particle spectrum and vacuum are concerned". Namely,  these theories should a have a unique maximally symmetric solution and a single massless spin-2 excitation and no other modes about this vacuum.  As a more stringent condition, we require that the theory has these two properties at every order in the curvature expansion. Therefore, according to this idea, as higher energies (or smaller distances ) are probed, gravity is deformed in such a way that 
no degrees of freedom in the gravity sector arise. The differences between  cosmological Einstein's gravity and its higher order cousins, that we have studied here, are simply encoded in the effective cosmological constant and the effective Newton's constant. In some sense, these two coupling constants run, as they do in quantum field theories, as higher energies are probed. Except, the running here is in terms of the curvature, not directly with the energy, and for the maximally symmetric background only the effective cosmological constant determines the value of the effective Newton's constant. 

In the original  works {\cite{tahsin2,tahsin3} Born-Infeld type extension of Einstein's gravity with only massless spin-2 modes about their unique vacua were constructed. The healthy Born-Infeld theory has 3 arbitrary parameters not fixed by the requirements that the vacuum is unique and the particle spectrum only consists of a massless spin-2 graviton. By a judicious choice of these parameters, one obtains a theory which is not of the Born-Infeld type but a cubic and quartic deformation of Einstein's gravity with specific combinations. The structure of the solutions of this  quartic theory has been the subject of this work as well as a detailed analysis of its particle and vacuum structure done in a more conventional way. It is rather remarkable that well-known singular solutions of Einstein's gravity are not exact solutions to this theory. For example, the Schwarzschild metric fails to be a solution. In the literature, some specific theories  that exclude the Schwarzschild metric were constructed before \cite{shortcuts,Des_Tek}, but these theories do not have the same particle spectrum as Einstein's gravity.  Furthermore, we have shown here that all Ricci flat solutions or Einsteinian solutions must satisfy a non-linear Klein-Gordon equation for the Kretschmann scalar (or the Gauss-Bonnet invariant which we utilized a lot ). This is a noteworthy result of its own: when curvature invariants  satisfy (nonlinear) wave-type equations of their own, singularities can be avoided on the basis of physical arguments.
We have also studied approximate spherically symmetric solutions far away from sources.  As the corrections to the Schwarzschild metric start at ${\cal O}(1/r^8)$, the solar system data is easily reproduced. A detailed study of the spherically symmetric solutions will be pursued elsewhere. Cosmological solutions and the inflation phase of the more general BI theory is studied in \cite{btekin} with promising results such as the existence of a quasi-de Sitter phase with enough number of $e$-foldings to solve the horizon problem.

\section{Acknowledgment}

 B.~T. is supported by the TUB\.{I}TAK grant 113F155 and thanks T.C. Sisman and I. Gullu for useful discussions.

\end{document}